\documentclass[prd,preprint,tightenlines,showpacs,aps,amsmath,amssymb,nofootinbib]{revtex4}

\usepackage{latexsym}
\usepackage{graphicx}
\usepackage{subfigure}

\usepackage{hyperref}
\usepackage{color}

\usepackage{amssymb}
\usepackage{bm}
\usepackage{natbib}

\begin{document}
\title{CMB Distortions from Superconducting Cosmic Strings}

\author{Hiroyuki Tashiro}
\email{Hiroyuki.Tashiro@asu.edu}
\author{Eray Sabancilar}
\email{Eray.Sabancilar@asu.edu}
\author{Tanmay Vachaspati}
\email{tvachasp@asu.edu}

\affiliation{
Physics Department, Arizona State University, Tempe, Arizona 85287, USA.
}


\begin{abstract}
We reconsider the effect of electromagnetic radiation from superconducting 
strings on cosmic microwave background (CMB) $\mu$- and $y$-distortions 
and derive present (COBE-FIRAS) and future (PIXIE) constraints on the 
string tension, $\mu_s$, and electric current, $I$. We show that absence 
of distortions of the CMB in PIXIE will impose strong constraints on 
$\mu_s$ and $I$, leaving the possibility of light strings 
($G\mu_s \lesssim 10^{-18}$) or relatively weak currents 
($I \lesssim 10~{\rm TeV}$).
\end{abstract}


\maketitle

\section{Introduction}
\label{sec:intro}

Phase transitions are key milestones in the thermal history of the universe. 
As the universe evolves and cools, fundamental symmetries are spontaneously 
broken and cosmological phase transitions occur. Therefore, the observation 
of phase-transition remnants can give us direct access to high energy particle 
physics and the very early universe.

Superconducting cosmic strings are one of a variety of topological defects that 
could be produced at phase transitions in the early universe 
\cite{2000csot.book.....V,Witten:1984eb}. As superconducting strings move 
through the cosmic magnetized plasma, they can develop and carry large 
currents, and oscillating loops of superconducting strings will emit 
copious amounts of electromagnetic radiation and particles mostly as bursts
\cite{Ostriker:1986xc,Vilenkin:1986zz}. This led to the idea that superconducting 
strings may be a candidate for the engine driving observed gamma ray 
bursts \cite{1987ApJ...316L..49B,1988ApJ...335..525P, Berezinsky:2001cp}, sources of cosmic ray bursts \cite{Berezinsky09} and radio transients \cite{Vachaspati:2008su,Cai:2011bi}.

In this paper we focus on the role of superconducting cosmic strings 
as sources that inject energy in the cosmic medium and cause spectral
distortions of the cosmic microwave background (CMB).
The measurement of the CMB spectral distortion is a good probe of
the thermal history of the universe and has been studied analytically 
and numerically in Refs.~\cite{Sunyaev:1980vz,1977NCimR...7..277D,
1982A&A...107...39D,1991A&A...246...49B,1991ApJ...379....1B, Hu:1992dc}.  
In the early universe ($z \gg 10^6$ where $z$ is the cosmic redshift), 
double Compton and Compton scatterings are very efficient, and any 
energy that is injected in photons into the cosmic medium is thermalized, 
and the cosmic radiation spectrum remains that of a blackbody.
However, the expansion of the universe makes these scatterings
less efficient with time and energy injected at epochs with $z<10^6$
produces CMB spectral distortions. That is, the spectrum departs
from a blackbody spectrum. Such distortions are commonly described by 
two parameters: the $\mu$ (chemical potential) distortion parameter, and 
the Compton $y$-parameter. Current constraints on these parameters have 
been obtained by COBE FIRAS and are:
$|\mu|<9 \times 10^{-5}$ and 
$y< 1.5 \times 10^{-5}$
\cite{Mather:1993ij,Fixsen:1996nj}.
The recently proposed future space mission called PIXIE has the 
potential to give dramatically tighter constraints on both
types of distortion, $|\mu| \sim5 \times 10^{-8}$ and
$y \sim 10^{-8}$ at the 5 $\sigma$ level \cite{Kogut:2011xw}.

There are several more conventional reasons for expecting CMB
distortions. The diffusion of density fluctuations before 
recombination, known as Silk damping \cite{Silk:1967kq}, is an 
energy injection source which produces CMB distortions 
\cite{1991MNRAS.248...52B,1991ApJ...371...14D,Hu:1994bz} at
the level of $\mu \sim 8 \times 10^{-9}$ 
\cite{Chluba:2011hw,Khatri:2011aj}. Other energy injection sources 
include massive unstable relic particles which decay during the 
thermalization epoch \cite{Hu:1993gc}, dissipation of primordial magnetic 
fields during the recombination epoch \cite{Jedamzik:1999bm}, and 
Hawking radiation from primordial black holes \cite{Tashiro:2008sf}.
An observation of CMB distortions will not by itself definitively 
point to a particular injection source, though upper limits on the 
distortions can be used to place constraints on models.

We evaluate both $\mu$- and $y$-distortions due to energy injected
from superconducting string loops. The injected energy depends 
on the current, $I$, carried by the strings, and on the string tension,
$\mu_s$. In general, the current arises due to the 
interaction of strings with ambient magnetic fields and need not be 
constant along a string, and may also vary among different parts
of the string network. However, we shall simplify our analysis by assuming 
the same constant current along all strings in the network. This 
simplification is expected to be accurate in the presence of 
primordial magnetic fields so that there is sufficient time for
the current to build up and saturate at its maximum possible value.

In this paper, we will obtain constraints in the two dimensional
parameter space given by the electric current on superconducting 
cosmic string loops and the string tension ($I-G\mu_s$ plane) due 
to present limits on CMB distortions. Early analyses placed a constraint 
on the fraction of electromagnetic to gravitational radiation from 
strings \cite{Ostriker:1986xc,Sanchez89,Sanchez90}. However, since both 
the electromagnetic and gravitational power depend on the string tension, 
as described in Sec.~\ref{sec:network}, those results cannot be directly 
used to produce a constraint plot in the $I-G\mu_s$ plane. Our analysis 
also differs from earlier work in details of the string network, and we 
are able to forecast constraints from future observation missions such 
as PIXIE. 

The organization of this paper is as follows. In Sec.~\ref{sec:network}, 
we discuss the cosmic string network properties and number density of loops, 
and then, derive the rate of electromagnetic energy density emitted from cusps 
of superconducting cosmic string loops. In Sec.~\ref{sec:cmbdist}, we calculate 
the spectral distortions of the CMB parametrized by chemical potential $\mu$ and 
Compton $y$-parameter due to cosmic strings, and obtain the corresponding 
constraints from COBE and PIXIE. Finally, in Sec.~\ref{sec:conclusions}, 
we summarize our findings.

Throughout this paper, we use parameters for a flat $\Lambda$CDM model: 
$h=0.7$ $(H_0=h \times 100 ~ {\rm km /s /Mpc})$, $\Omega_b=0.05$  and 
$\Omega_m=0.26$. Note also that $1+z=\sqrt{t^{\prime}/t}$ in the radiation 
dominant epoch and $1+z=(1+z_{\rm eq}) (t_{\rm eq}/t)^{2/3}$ in the matter 
dominant epoch, where $t^{\prime} = (2 \sqrt{\Omega_r} H_0)^{-1}$ and 
$z_{\rm eq}=\Omega_m / \Omega_r$ with $h^2 \Omega_r=4.18 \times 10^{-5}$.
We also adopt natural units, $\hbar = c =1$, and set the Boltzman constant 
to unity, $k_B=1$.


\section{String network and radiation}
\label{sec:network}

A superconducting string loop emits electromagnetic radiation at frequency
harmonics defined by its inverse length. The emitted power is dominated by
the highest frequency and is cut off by the finite thickness of the string.
The total power emitted in photons from loops with cusps is 
\cite{Vilenkin:1986zz}
\begin{equation}
P_\gamma = \Gamma_\gamma I \sqrt{\mu_s},
\label{eq:Pgamma}
\end{equation}
where $I$ is the current on the string (assumed constant), $\mu_s$ is
the string tension, and $\Gamma_\gamma \sim 10$ is a numerical coefficient
that depends on the shape of the loop. The string network also contains
a similar number of loops without cusps which emit much less power, $P \sim I^{2}$, in 
electromagnetic radiation than loops with cusps. Therefore, the contribution of cuspless loops can be ignored.

The loop also emits gravitational radiation with power \cite{Vachaspati:1984gt}
\begin{equation}
P_g = \Gamma_g G \mu_s^2,
\label{eq:Pg}
\end{equation}
where $\Gamma_g \sim 100$. Therefore, for every $\mu_s$, there is a critical
current 
\begin{equation}
I_* = \frac{\Gamma_g G\mu_s^{3/2}}{\Gamma_\gamma},
\label{eq:I*}
\end{equation} 
and for $I > I_*$ electromagnetic radiation dominates and determines 
the lifetime of the loop, while for $I< I_*$ gravitational losses
are more important. Hence, we can write the lifetime of a string loop
of length $L$ as
\begin{equation}
\tau = \frac{L}{\Gamma G\mu_s},
\end{equation}
where 
\begin{eqnarray}
\Gamma &=& \Gamma_g \ , \ \ I < I_*, \nonumber \\
\Gamma &=& \frac{\Gamma_\gamma I}{G\mu_s^{3/2}}
       = \Gamma_g \frac{I}{I_*} \ , \ \ I > I_* \ .
\label{eq:gammadefn}
\end{eqnarray}
If a loop is born with length $L_i$, its length changes with time as
\begin{equation}
L = L_i - \Gamma G\mu_s (t-t_i).
\end{equation}
Assuming slow decay, we take $t \gg t_i$ and hence
\begin{equation}
L_i \approx L + \Gamma G\mu_s t.
\label{Lvst}
\end{equation}

Analytical studies \cite{Rocha:2007ni,Polchinski07,Dubath08,Vanchurin11,Lorenz:2010sm} and simulations \cite{Bennett90,Allen90,Hindmarsh97,Martins06,Ringeval07,Vanchurin06,
Olum07,Shlaer10,Shlaer11} all yield a consistent picture for large cosmic string 
loops but differing results for small loops.
The results can be summarized during the cosmological radiation dominated 
epoch ($t < t_{\rm eq}$), by giving the number density of loops of 
length between $L_i$ and $L_i+dL_i$, 
\begin{equation}
dn(L_i,t) = \kappa \frac{dL_i}{t^{4-p} L_i^{p}}.
\label{eq:dnLi}
\end{equation}
The overall normalization factor, $\kappa$, will be assumed to be $\sim 1$. 
In our calculation, we shall take the exponent $p=2.5$, though somewhat
different values are suggested in other studies, 
{\it e.g.}, $p=2.6$ and $\kappa \sim 0.1$ in \cite{Lorenz:2010sm}.
Our final constraints are not affected significantly by such a slight
increase in the value of $p$, especially because the increased effect
from small loops is compensated by the smaller value of $\kappa$.

Inserting Eq.~(\ref{Lvst}) in (\ref{eq:dnLi}) gives
\begin{equation}
dn(L,t) = \kappa \frac{dL}{t^{3/2} (L + \Gamma G\mu_s t)^{5/2}}\ , 
           \ \ t < t_{\rm eq},
\end{equation}
as the number density of loops of length $L$ at cosmic time $t$.

Similarly, during the cosmological matter dominated epoch the number
density of loops is
\begin{equation}
dn(L,t) = \frac{\kappa C_L dL}{t^2 (L+\Gamma G\mu_s t)^2}\ , 
              \ \ t > t_{\rm eq},
\label{eq:ndensity}
\end{equation}
where
\begin{equation}
C_L \equiv  1 + \sqrt{\frac{t_{\rm eq}}{L+\Gamma G\mu_s t}}\ .
\end{equation}
The second term takes into account the loops from the radiation 
dominated epoch that survive into the matter dominated epoch. 

%
%
%

The energy injection rate into photons from cosmic strings is found by
multiplying Eq.~(\ref{eq:Pgamma}) by the number density of loops in
Eq.~(\ref{eq:ndensity}), and integrating over loop length
\begin{equation}
{d Q\over dt} =  \Gamma_\gamma I \sqrt{\mu_s} \int_0^\infty dn(L,t) \ .
\end{equation}

\begin{figure}
   \includegraphics[width=100mm]{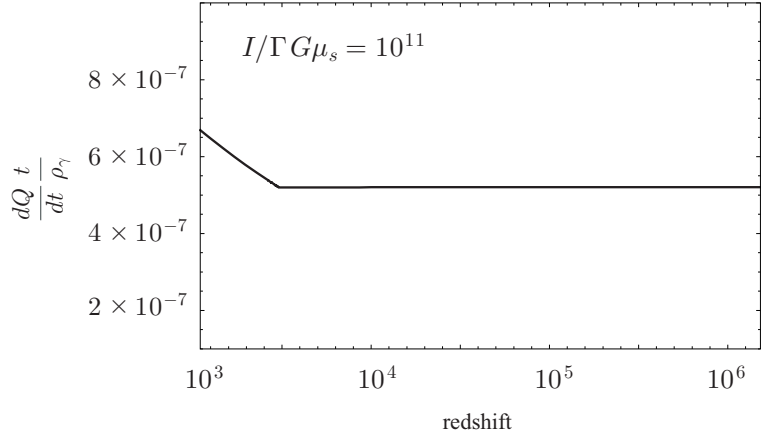}
\caption{The redshift evolution of ${dQ/dt}$. 
We take $I/ \Gamma G\mu_s = 10^{11}~$GeV and assume 
$\Gamma G\mu_s t_0 \ll t_{\rm eq}$ (see Eq.~(\ref{eq:dqdtMD})).}
  \label{fig:energy-gmu}
\end{figure}

We can write $dQ/dt$ in the radiation and the matter dominated 
epochs as
\begin{equation}
\frac{dQ}{dt} = \frac{2\kappa \Gamma_\gamma}{3(\Gamma G\mu_s )^{3/2}}
                \frac{I\sqrt{\mu_s}}{t^3} \ , \ \ t < t_{\rm eq},
\label{eq:dqdtRD}
\end{equation}
and
\begin{eqnarray}
\frac{dQ}{dt} &=& \frac{\kappa \Gamma_\gamma}{\Gamma G\mu_s}
                \frac{I\sqrt{\mu_s}}{t^3} 
  \left [ 1 + \frac{2}{3}\sqrt{\frac{t_{\rm eq}}{\Gamma G\mu_s t}} ~\right ]
  \ , \ \ t > t_{\rm eq}, \nonumber \\
&\simeq& \frac{2\kappa \Gamma_\gamma}{3(\Gamma G\mu_s )^{3/2}}
     \frac{I\sqrt{\mu_s}}{t^3} \sqrt{\frac{t_{\rm eq}}{t}}, 
\label{eq:dqdtMD}
\end{eqnarray}
where in the second line, we have restricted attention to strings
such that $\Gamma G\mu_s t_0 \ll t_{\rm eq}$ --- $t_0$ being the present
cosmological epoch --- a condition that is satisfied
in a large part of the allowed range of string parameters.

%



\section{CMB distortions due to cosmic strings}
\label{sec:cmbdist}

In the early universe ($z>10^6$ where $z$ denotes cosmic redshift), we expect that energy 
injected into the cosmological medium will be thermalized by photon-electron interactions, 
i.e., by Compton and double Compton scatterings. As a result, the photon distribution in 
the early universe maintains its blackbody spectrum. However, the energy injection from 
cosmic strings mainly consists of very high energy photons ($\omega \sim \sqrt{\mu_s} \gg m_e$) 
and the optical depth of such high energy photons for Compton and double Compton scatterings 
is not high, because the scattering cross-sections are suppressed by the photon energy. Then, 
thermalization proceeds in two steps. First, the high energy photons lose their energy 
quickly via photon-photon scattering or photo pair production
(see Ref.~\cite{1989ApJ...344..551Z}). Once the energy of the photons is reduced by
these processes, they can be thermalized by Compton and double Compton scatterings and 
the photons again achieve a blackbody spectrum.

At lower redshifts ($z < 10^6$), Compton and double Compton scatterings decouple and 
the injected photons can no longer be thermalized efficiently. Accordingly, energy 
injection produces distortions in the blackbody spectrum of the CMB.

First, the decoupling of double Compton scattering takes place at
$z \sim 10^6$. As a consequence, photon number is conserved
for $z < 10^6$, and only the energy among the photons can be
re-distributed. This is insufficient to establish a blackbody
spectrum for the photons. However, the injected photons are still 
thermalized by Compton scattering, and the CMB spectrum in this thermal equilibrium 
state is described by the Bose-Einstein distribution with a chemical potential $\mu$.

Thermalization due to Compton scattering also becomes inefficient
at $z \sim 10^5$, when the time scale of the Compton scattering
process becomes longer than the Hubble time. The energy injection after 
Compton decoupling produces a distortion which is parameterized by the 
Compton $y$-parameter.




\subsection{$\mu$-distortion}
\label{sec:mudist}

The time evolution of the $\mu$-distortion of the CMB spectrum due to energy injection 
is given by \cite{Hu:1992dc}
\begin{equation}
\frac{d\mu}{dt} = -\frac{\mu}{t_{DC}(z)} + \frac{1.4}{\rho_\gamma} \frac{dQ}{dt}\ .
\label{eq:muevoln}
\end{equation}
Here $\rho_\gamma$ is the photon energy density, $t_{DC}$ is the time scale for double 
Compton scattering
\begin{equation}
t_{DC} = 2.06\times 10^{33} \left (1-\frac{Y_p}{2} \right )^{-1}
                                       (\Omega_bh^2)^{-1}z^{-9/2}~ {\rm s},
\end{equation}
where $Y_p$ is the primordial helium mass fraction.  

\begin{figure}
   \includegraphics[width=100mm]{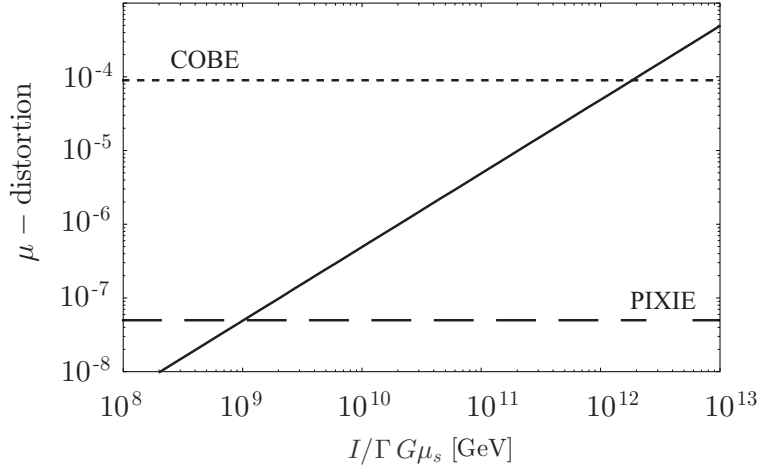}
  \caption{The $\mu$-distortion as a function of $I/ \Gamma G \mu_s$.
The dotted and dashed lines represent COBE FIRAS limit and the detection 
limit by PIXIE in its current design.} 
  \label{fig:mu-distortion}
\end{figure}

\begin{figure}
   \includegraphics[width=100mm]{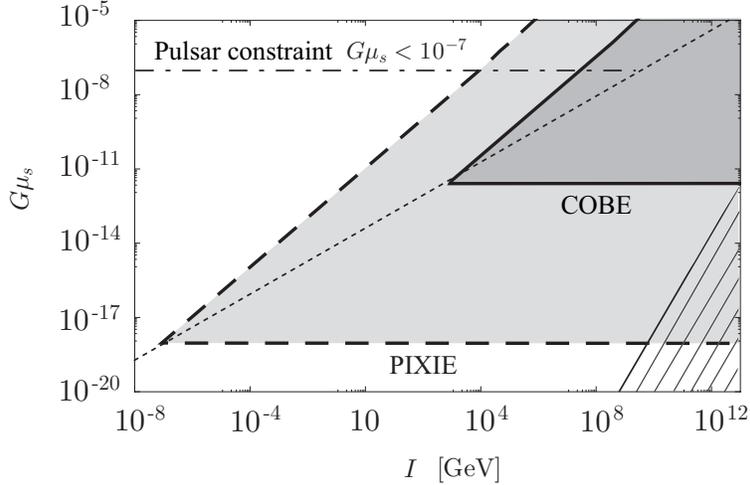}
\caption{The constraint from $\mu$-distortion on $I$--$G \mu_s$ plane. The 
dark shaded area is ruled out by COBE constraint on $\mu$-distortion. If there
is no detection of $\mu$-distortion by PIXIE, the lightly shaded region within 
the thick 
dashed line will be ruled out. The region above the thin dashed line is 
where gravitational radiation dominates over electromagnetic radiation, i.e.,
$I < I_*$ (see Eq.~(\ref{eq:I*})). The hatched region is excluded because 
$I > \sqrt{\mu_s}$, and exceeds the saturation value of the current on superconducting strings. Also, millisecond pulsar observations constrain $G\mu_s \lesssim 10^{-7}$
as shown by the dot-dashed line.}
  \label{fig:mu-gmuI}
\end{figure}

As explained above, the $\mu-$distortion is only produced in a redshift range
$z_1\sim 10^6$ to $z_2 \sim 10^5$, when double Compton scattering is inefficient,
but Compton scattering is still operative. Then, the
solution to Eq.~(\ref{eq:muevoln}) is
\begin{equation}
\mu = 1.4\int_{t(z_1)}^{t(z_2)}dt\frac{dQ/dt}{\rho_{\gamma}}
\exp[-(z(t)/z_{DC})^{5/2}] = 1.4\int_{z_1}^{z_2}dz\frac{dQ/dz}{\rho_{\gamma}} \exp[{-(z/z_{DC})^{5/2}}], 
\label{eq:evomu}
\end{equation}
where
\begin{equation}
z_{DC} = 1.97 \times 10^6\left[1-\frac{1}{2}\left(\frac{Y_p}{0.24}\right)\right]^{-2/5}
              \left(\frac{\Omega_bh^2}{0.0224}\right)^{-2/5},
\end{equation}
Performing the integration in Eq.~(\ref{eq:evomu}) with Eq~(\ref{eq:dqdtRD}),
we find
\begin{equation}
\mu =4.6 \times 10^{-6} \left({I/ \Gamma G \mu_s \over 10^{11} ~{\rm GeV}} \right),
\end{equation}
which is plotted in Fig.~\ref{fig:mu-distortion} (the result depends only very weakly
on the integration limits $z_1$ and $z_2$).
According to the COBE constraint \cite{Mather:1993ij,Fixsen:1996nj}, we
obtain 
\begin{equation}
{I \over \Gamma G \mu_s} < 1.95 \times 10^{12} ~{\rm GeV} \ , \ \ {\rm (COBE)},
\label{eq:igmu-cobe}
\end{equation}
and the predicted constraint from PIXIE is more severe \cite{Kogut:2011xw},
\begin{equation}
{ I \over  \Gamma G \mu_s} < 1.08 \times 10^{9} ~{\rm GeV}\ , \ \ {\rm (PIXIE)}.
\label{eq:igmu-pixie}
\end{equation}
We plot these constraints in the $I$--$G\mu_s$ plane in Fig.~\ref{fig:mu-gmuI}. 

In Fig.~\ref{fig:mu-gmuI}, the region above the short-dashed line 
is where gravitational radiation losses dominate the electromagnetic 
radiation, i.e., $I<I_*$, where $I_*$ is defined in Eq.~(\ref{eq:I*}).
In this region, $\Gamma \sim 100$ from Eq.~(\ref{eq:gammadefn}). Then,
Eqs.~(\ref{eq:igmu-cobe}) and (\ref{eq:igmu-pixie}) give the constraints,
\begin{equation}
{I \over G \mu_s} < 1.95 \times 10^{14} ~{\rm GeV}\ , \ \ {\rm (COBE)},
\label{eq:igmu-cobe-2}
\end{equation}
\begin{equation}
{ I \over  G \mu_s} < 1.08 \times 10^{11} ~{\rm GeV}\ , \ \ {\rm (PIXIE)}.
\label{eq:igmu-pixie-2}
\end{equation}
In the region below the short-dashed line in Fig.~\ref{fig:mu-gmuI}, 
electromagnetic radiation dominates over gravitational radiation. 
As a result, in this region, the constraints from $\mu$-distortion
represented by Eqs.~(\ref{eq:igmu-cobe}) and (\ref{eq:igmu-pixie}) can be
written using Eq.~(\ref{eq:gammadefn}),
\begin{equation}
 G \mu_s < 2.5 \times 10^{-12},
\quad {\rm (COBE)},
\end{equation}
\begin{equation}
 G \mu_s < 7.9 \times 10^{-19},
\quad {\rm (PIXIE)}.
\end{equation}
Note that the constraint is independent of the current provided
$I > I_*$ holds. 




\subsection{Compton $y$-distortion}
\label{sec:ydist}

For $z < 10^5$, the injected energy is no longer thermalized by 
Compton scattering. Instead, the injected energy heats up electrons, 
which then scatter the CMB photons by the inverse Compton process,
leading to $y$-distortions of the CMB.
 
The Compton $y$-parameter is given by \cite{Sunyaev:1980vz}
\begin{equation}
y=\int_{t_{\rm freeze}}^{t_0} dt \frac{T_e-T}{m_e} n_e \sigma_T ,
\label{eq:y-para-def}
\end{equation}
where $T_e$ is the electron temperature, $T$ is the temperature
of the cosmic background radiation,
$n_e$ is the number density of free electrons, $\sigma_T$ is the
Thomson scattering cross section and $t_0$ is the present time.  The
time $t_{\rm freeze}$ represents the freeze out time of thermalization, which we set it to be $z\sim 10^5$.

The evolution of the electron temperature $T_e$ with injected photon
energy is written as
\begin{equation}
%
n_e {d \over dt} T_e  = {n_e \sigma_T \over 3} 
             \int {\omega -4  T_e \over m_e } \omega f_\omega d \omega 
     -{4\over 3} { n_e\sigma_T   \over   m_e } \rho_{\gamma} (T_e-T) 
     -2{\dot a \over a}  T_e
n_e,
\label{eq:etemp}
\end{equation}
where $f_\omega$ is the spectrum of photons injected by time $t$ --- 
in other words, $f_\omega$ is the spectrum of all photons minus the 
spectrum of blackbody photons. 
The first term on the right hand side (rhs) of Eq.~(\ref{eq:etemp}) 
describes Compton heating of electrons by injected photons; the 
second term describes the Compton cooling of electrons by photons; 
the third describes cooling due to cosmic expansion. 

The evolution of the spectrum $f_\omega$ is obtained from the equation,
\begin{equation}
{\partial f_\omega \over \partial t} =
{\omega -4  T_e \over m_e } \omega  {\partial f_\omega  \over \partial
\omega } n_e \sigma_T
+{2 \omega -4  T_e \over m_e } f_\omega  n_e \sigma_T 
+ {\dot a \over a} \omega  {\partial f_\omega  \over \partial \omega}
-2 {\dot a \over a}  f_\omega
+ \delta f_\omega,
\label{eq:pnum}
\end{equation}
where $\delta f_\omega$ is the injected number of photons with frequency 
$\omega$ per unit time.
The first two terms on the rhs of Eq.~(\ref{eq:pnum}) describe Compton 
cooling of injected photons, and the third and forth terms describe
cooling due to Hubble expansion.

In order to analytically evaluate the electron temperature, we note
that the last term in Eq.~(\ref{eq:etemp}) is suppressed by the
inverse cosmic time, which is much larger than the microphysical
time involved in Compton processes. So we ignore the Hubble expansion
term and assume the quasi-steady state condition: $dT_e /dt =0$. 
Then, the electron temperature is
\begin{equation}
T_e -T ={m_e  \over 4 \rho_\gamma} 
\int ^{\infty} _{0} d\omega
            {\omega -4  T_e \over m_e } \omega  f_\omega. 
\label{eq:etemp2}
\end{equation}
The term on the rhs can be found by integrating Eq.~(\ref{eq:pnum})
over the photon energy
\begin{equation}
{\partial \over \partial t} \int^{\infty}_{0} d\omega  \omega  f_\omega 
=-\int^{\infty}_{0} d\omega  {\omega -4  T_e \over m_e } \omega
f_\omega  n_e \sigma_T 
+{ d Q \over dt},
\label{eq:penergy}
\end{equation}
where we ignore cosmic expansion again because the time scale of 
Thomson scattering is much shorter than the cosmological time. 
Also, the total injected energy rate by cosmic strings is
\begin{equation}
{d Q \over dt} = \int d \omega  \omega ~ \delta f_\omega \ .
\end{equation}
The first term on the rhs of Eq.~(\ref{eq:penergy}) describes the 
energy loss rate from the photons due to Compton cooling.
We express this term as $\partial {\cal E}_\gamma/\partial t$.
Note that $f_\omega$ is the spectral distribution of photons
minus the blackbody distribution, and ${\cal E}_\gamma$ is
also the energy of photons that are in the spectral deviation 
from blackbody.

Then, from Eq.~(\ref{eq:etemp2}), we can rewrite the electron
temperature in terms of $E_{\rm loss}$ as
\begin{equation}
T_e -T = \frac{m_e}{4\rho_\gamma n_e \sigma_T} 
 \left [ \frac{dQ}{dt} - \frac{\partial {\cal E}_\gamma}{\partial t} \right ].
\label{eq:TeTQtEt}
\end{equation}

\begin{figure}
   \includegraphics[width=100mm]{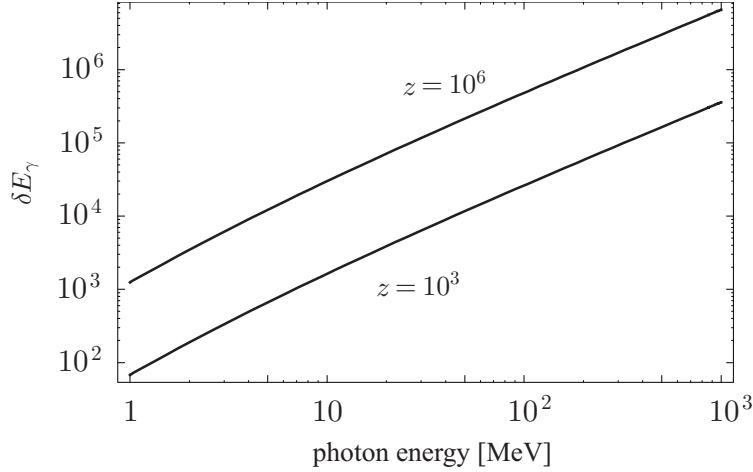}
  \caption{The energy loss from photons in a Hubble time due to
Compton scattering as a function of injected photon energy.
The curves are for $z=10^6$ and $10^3$.}
  \label{fig:energyloss}
\end{figure}

We will now argue that the rate of change of ${\cal E}_\gamma$ is of 
order the Hubble expansion rate and can be ignored in our quasi-steady
state treatment. Basically, the idea is that the energy of high
frequency photons injected by strings is transferred very efficiently 
to electrons by Compton scattering. 
A photon with frequency $E_\gamma$ loses energy 
$E_\gamma (E_\gamma -4T_e )/m_e $ per Compton
scattering.  Hence, the energy loss of a photon 
with high initial energy $E_{\gamma0} \gg T_e$ 
within a Hubble time is approximated as
\begin{equation}
\delta E_\gamma \simeq {E_{\gamma0}^2 \over m_e} {n_e \sigma_T  \over H}.
\end{equation}
Fig.~\ref{fig:energyloss} shows that $\delta E_\gamma \gg E_{\gamma 0}$ 
at $z=10^6$ and $10^3$ for high energy photons and  
it is clear that the injected photon energy is fully transferred into 
electrons well within a Hubble time. So, the energy ${\cal E}_\gamma$ 
only varies on a cosmological time scale, and its time derivative
can be ignored in the quasi-steady state approximation.
Therefore, we can drop the last term in Eq.~(\ref{eq:TeTQtEt}) and obtain
\begin{equation}
T_e -T \approx \frac{m_e}{4\rho_\gamma n_e \sigma_T} \frac{dQ}{dt}, 
\label{eq:TeTQt}
\end{equation}
which, from Eq.~(\ref{eq:y-para-def}), leads to
\begin{equation}
y=\frac{1}{4} \int^{t(z_{\rm rec})}_{t(z_{\rm freeze})} dt 
          \frac{1}{\rho_r} \frac{dQ}{dt} \ ,
\label{eq:y-dis}
\end{equation}
where the upper bound of the integration $t(z_{\rm rec})$ is the
recombination epoch, which is introduced since the injected energy
does not transfer into the background electrons once the optical depth 
becomes very low after recombination.

\begin{figure}
   \includegraphics[width=100mm]{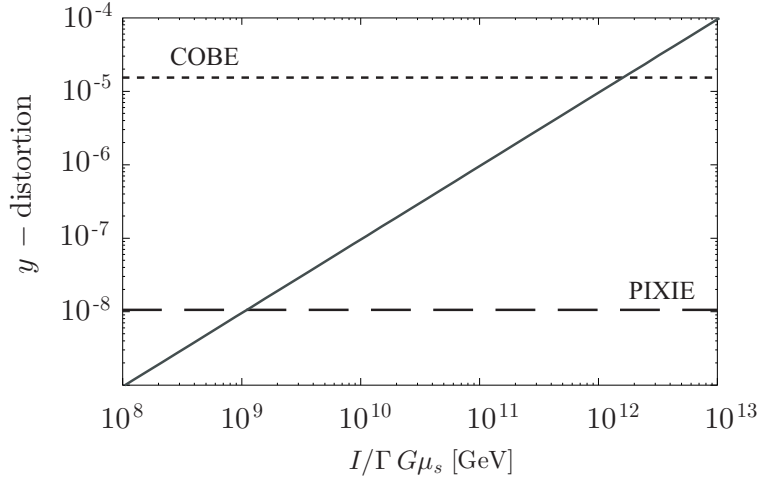}
  \caption{The $y$-distortion as a function of $I/\Gamma G \mu_s$.
The dotted and dashed lines represent 
COBE FIRAS limit and the current detection limit by PIXIE,
respectively. 
}
  \label{fig:y-distortion}
\end{figure}

We now calculate the integral in Eq.~(\ref{eq:y-dis}) with 
Eq.~(\ref{eq:dqdtRD}) and Eq.~(\ref{eq:dqdtMD}), and
plot the result in Fig.~\ref{fig:y-distortion}. The fit is
\begin{equation}
y =9.53 \times 10^{-7} \left( 
            {I/\Gamma G \mu_s \over 10^{11} ~{\rm GeV}} \right).
\end{equation}
The corresponding constraint from COBE yields
\begin{equation}
\frac{I}{\Gamma G \mu_s} < 1.57 \times 10^{12} ~{\rm GeV} \ , \ \ {\rm (COBE)}
\label{eq:igmu-cobe-y}
\end{equation}
and PIXIE will be able to constrain up to 
\begin{equation}
\frac{I}{\Gamma G \mu_s} < 1.09 \times 10^{9} ~{\rm GeV} \ , \ \ {\rm (PIXIE)}.
\label{eq:igmu-pixie-y}
\end{equation}
These constraints are very similar to those obtained from $\mu$-distortion 
in Eqs.~(\ref{eq:igmu-cobe-2}) and (\ref{eq:igmu-pixie-2}).

\section{Conclusions}
\label{sec:conclusions}

We studied the effect of electromagnetic radiation from superconducting 
cosmic string loops on CMB spectral distortions, and obtained constraints 
on the parameter space of string tension $G \mu_s$ and the current $I$. 
Earlier studies by Refs.~\cite{Sanchez89,Sanchez90} to constrain 
superconducting cosmic string parameters from CMB spectral distortions 
assumed that the power going into electromagnetic radiation, $P_{\gamma}$, 
is a $\mu_s$ independent, small, constant fraction of the power going into 
gravitational radiation, $P_{g}$, and hence, the loop lifetime is 
determined by gravitational radiation. In Sec.~\ref{sec:network}, we 
explained that the electromagnetic power depends on the current in the 
string [see Eq.~(\ref{eq:Pgamma})], hence, it is not simply a fraction 
of $P_{g}$. Besides, since $P_{\gamma}$ depends on the current, $I$, at 
some value of the current given by Eq.~(\ref{eq:I*}), the electromagnetic 
radiation becomes the dominant energy loss mechanism, and the lifetime 
of the loops is determined by $P_{\gamma}$. 

We made some simplifying assumptions in this paper. First of all, we 
assumed that the cosmic network characteristics for the superconducting 
strings is the same as ordinary ones with no current, as always assumed 
in cosmic string simulations. We do not think that the effect of the 
current will be very significant in the accuracy of our order of magnitude 
estimates. Another simplifying assumption was that cosmic string cusps 
produce homogeneous CMB distortions. Therefore, we assumed that the beamed 
radiation from cusps are quickly isotropized since we focus on very early 
epochs $z < z_{rec} \sim 1100$, where the injected photons quickly 
thermalize \cite{1989ApJ...344..551Z}. On the other hand, if the radiation 
from cusps is not isotropized efficiently, the CMB distortions will depend
on the direction of observation. 

In Sec.~\ref{sec:cmbdist}, we showed that both $\mu$- and $y$-distortions 
give comparable constraints on the parameter space. COBE-FIRAS measurement 
of no spectral distortion of the CMB places upper bounds on the 
distortion parameters, $|\mu|<9 \times 10^{-5}$ and 
$y< 1.5 \times 10^{-5}$ \cite{Mather:1993ij,Fixsen:1996nj}. On the other 
hand, the proposed future space mission PIXIE can constrain them up to, 
$|\mu| \sim5 \times 10^{-8}$ and $y \sim 10^{-8}$ at the 5 $\sigma$ 
level \cite{Kogut:2011xw}. The corresponding constraints from COBE 
and PIXIE on string parameters for $\mu$ distortion are relatively 
given by 
\begin{eqnarray}
{I \over G \mu_s} < 1.95 \times 10^{14} ~{\rm GeV},~~ {\rm(COBE)},\\
{ I \over  G \mu_s} < 1.08 \times 10^{11} ~{\rm GeV},~~{\rm(PIXIE)},
\end{eqnarray}
where the loop lifetime is determined by gravitational energy losses, 
$I < I_{*}$. In the opposite regime, $I> I_{*}$, we obtained
\begin{eqnarray} 
 G \mu_s < 2.5 \times 10^{-12},~~{\rm(COBE)},\\
 G \mu_s < 7.9 \times 10^{-19},~~{\rm(PIXIE)}.
\end{eqnarray}
These constraints are summarized in Fig.~\ref{fig:mu-gmuI}. We have also calculated the CMB $y-$distortion due to superconducting strings. These lead to constraints that are similar in magnitude to those from the $\mu-$distortion and are shown in  Fig.~\ref{fig:y-distortion}.

If PIXIE does not detect suitable distortions, only light superconducting strings with modest currents (up to $\sim 10^8 ~{\rm GeV}$), or somewhat heavier strings but with small currents ($\lesssim 10^4 ~{\rm GeV}$) will be allowed. Of course, there is a possibility that CMB distortions will be detected, in which case, one needs to look at other distinguishing signatures from superconducting cosmic strings such as neutrino bursts \cite{Berezinsky09} and radio transients \cite{Vachaspati:2008su,Cai:2011bi}.


\begin{acknowledgments}
This work was supported by the DOE and by NSF grant PHY-0854827 at Arizona State University.
\end{acknowledgments}



%
%
%
%
%
%
%
%
%

\end{document}